\documentclass[aps,twocolumn,groupedaddress,showpacs,showkeys]{revtex4}
\usepackage{graphicx}
\begin{document}
\title{Imperfect Homoclinic Bifurcations}
\author{Paul Glendinning}
\affiliation{Department of Mathematics, UMIST, P.O. Box 88, 
Manchester M60 1QD, UK}
\author{Jan Abshagen \& Tom Mullin}
\affiliation{Manchester Center for Nonlinear Dynamics, University of
Manchester, Oxford Road, Manchester M13 9PL, UK}
\date{\today}
\begin{abstract}
Experimental observations of an almost symmetric electronic circuit show
complicated sequences of bifurcations. These results are discussed in the
light of a theory of imperfect global bifurcations. It is shown that 
much of the dynamics observed in the circuit can be 
understood by reference to imperfect homoclinic bifurcations without 
constructing an explicit mathematical model of the system. 
\end{abstract}
\maketitle

\section{Introduction}
\label{intro}
The r\^{o}le of symmetries in determining the behaviour of nonlinear physical 
systems can be crucial. Reflection (or ${\bf Z}_2$) symmetry is relevant
to a wide range of experiments, and in such a system a pair of
stable solutions may be created by a supercritical pitchfork bifurcation as a 
parameter is varied. These new states break the original symmetry, but are
symmetric images of each other. Of course, perfect symmetry is
never achievable in any physical system so in practice the bifurcation 
may become disconnected  having one branch which varies 
monotonically with the parameter and a second which arises by a saddle--node 
bifurcation. This is most easily 
modelled by adding an imperfection term as a constant in the model normal form 
and this appears to work well in describing the local bifurcation structure. 
However, a physical system will typically contain many sources for this 
imperfection and some of them may be high--dimensional in nature. Therefore, 
it is reasonable to ask whether a model with a single imperfection term provides a good 
representation of the system far from the bifurcation point. Specifically, we
are interested here in the effects of this local modelling 
on the global dynamics which result from homoclinic bifurcations.

Our investigation is concerned with a class of global bifurcations
involving homoclinic orbits, i.e. orbits which tend to a stationary point
of the model flow in both forwards and backwards time. Typically, the existence of a
homoclinic orbit is not a persistent property of a differential equation, but they
occur on lines in two-parameter families (technically, they are codimension
one bifurcations). In the absence of symmetry, the net effect of such bifurcations 
is to create or destroy a periodic orbit, whose period tend to infinity at the 
bifurcation point. This may happen in one of two ways: one-sided or two-sided. In the
one-sided case, the orbit apporaches the bifurcation point from one side of
the bifurcation point as its period tends to infinity. In the two-sided
case, such as the Shil'nikov case \cite{Shil}, the locus of the orbit in parameter 
space oscillates about the bifurcation value creating the so--called  `Shil'nikov wiggle'
as the period of the  orbit tends to infinity. Moreover,
there are period-doubling and reverse period-doubling bifurcations of
the orbit together with more complicated homoclinic bifucations. This sequence of
events has been reported previously \cite{He} in an experimental and theoretical study
of a modified van der Pol oscillator, and in a wide variety of other 
experiments including Taylor--Couette flows \cite{mull,pfist}, optics, 
\cite{arrec,herr}, chemical oscillators \cite{haus, arne} and liquid crystal 
flows \cite{peac}. 

In the presence of simple symmetries, homoclinic bifurcations may involve
two or more homoclinic orbits. In the simplest cases
the net effect is to destroy a pair of
periodic orbits which are the image of each other under the symmetry and create
a single symmetric branch of periodic orbits. These symmetric periodic orbits
cannot undergo period-doubling bifurcations in the two sided case.  The 
period-doubling and
reverse period-doubling bifurcations on branches of the symmetric orbit 
are replaced by an initial symmetry-breaking (or reverse
symmetry-breaking) bifurcation. The asymmetric orbits created in this way may,
of course, be involved in period-doubling bifurcations. This distinction will
be useful in the interpretation of the bifurcations observed below.

Whilst the effect of small symmetry-breaking terms on the bifurcations of
stationary solutions has a long history (the imperfection theory
of Golubitsky and Schaeffer \cite{GSh1,GSh2,Sch}) there appears to have been
no systematic attempt to describe the equivalent modifications of global
bifurcations (although see \cite{C,gl} for a special case). Our aim here is to
provide the foundations for such an approach. We reconsider the experimental
electronic oscillator \cite{He} which exhibits a variety of almost symmetric
global bifurcations and show how many features observed in the
experiments may be explained by reinterpreting some results on
codimension two homoclinic bifurcations so as to obtain a general
imperfection theory for homoclinic bifurcations.
These results necessarily involve non-stationary solutions, and so are
likely to be applicable and observable in many more interesting situtations.

The experiments were carried out using a van der Pol oscillator.
The bifurcation structure of this system has been investigated in detail
previously \cite{He} but with the implicit assumption of symmetry. It is the aim of the
present study to investigate the global dynamics of the circuit and relate the
observations to modern ideas on gluing bifurcations where the mathematical
abstraction of perfect symmetry is relaxed.

\section{Experiment I}
\label{experiment1}
\subsection{The electronic oscillator}
\label{circuit}
The experimental study was performed using a van der Pol oscillator circuit,
the details of which are given in Healey et al.~\cite{He}. It comprises an autonomous LCR
oscillator with two nonlinear conductances in the feedback circuit. Precise variation of the 
two parameters
which control the behaviour of the system was provided by switchable decades
resistance boxes. By this means determination of the bifurcation
structure to a relative accuracy of  better than $0.1\%$ was possible. The two parameters are
denoted by $\alpha_{1},\beta_{1}$ and they are nondimensionalised forms of the resistances
$R_{1},R_{2}$ which control the nonlinear elements. Details of the nondimensionalisation are given
in Healey et al.~\cite{He}.

The principle set of observations were made using an oscilloscope. Steady 
bifurcations were observed as changes in the level of the d.c.~output.
On the other hand dynamical states were best monitored as Lissajous figures formed from
a combination of signals measured over the nonlinear elements. In this way, limit cycles,
period doubling sequences, chaos etc. were readily displayed. Time-series were
also recorded and stored on a computer  via a 12-bit A/D for further processing. This
included phase portrait analysis using the method of delay coordinates.

The inductor used in the present circuit is 1.5269H compared with 1.78H used by 
Healey et al.~\cite{He}. This causes a shift of the bifurcation points
relative to those previously reported, though the 
bifurcation structure remains qualitatively the same. The imperfections in the
circuit are tiny and the resulting disconnections are equally small. They arise
from a variety of sources but we will refer to them throughout as a single
imperfection.

\subsection{Bifurcation set}

\begin{figure}
\includegraphics[width=8cm]{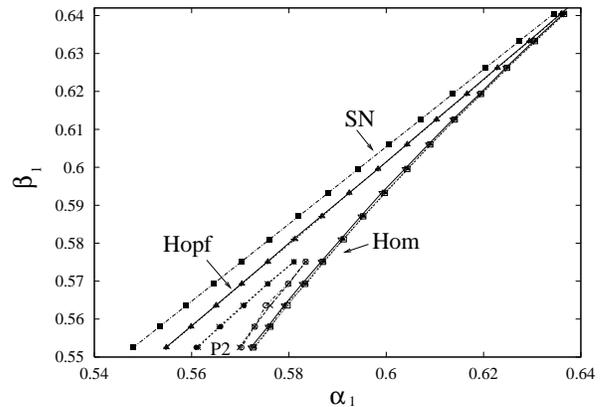}
\caption{Experimental bifurcation set in the $\alpha_{1},\beta_{1}$ plane. SN
denotes the path of saddle-node bifurcations, `Hopf' the Hopf bifurcations to
simple oscillations and `Hom' the gluing bifurcations. The paramater region
denoted by `P2' is where forward and reverse period doubling is observed on the
asymmetric orbits.}
\label{fig1}
\end{figure}

The stability diagram for the electronic circuit is shown in Figure \ref{fig1}. 
The overall structure shows lines of steady and dynamic bifurcations all meeting 
at the top right hand corner of the figure which is a codimension--2 point. 
The dynamic bifurcations (Hopf and homoclinic) are pairs of lines superposed and 
separated by the imperfections in the circuit. This effect is very small and 
cannot be resolved on the scale of the figure but, as we will show below, 
it has a significant effect on the global dynamics.

In the parameter range of interest, a perfectly symmetric system would have
a trivial zero volts fixed point which would lose stability to
a pair of non--zero d.c.~states at a supercritical
pitchfork bifurcation. As expected, in the experiment
we see that this bifurcation is disconnected to form a 
continuously connected state and a separate
solution branch which is terminated at its lower end by a saddle--node 
bifurcation denoted by SN in Figure \ref{fig1}. The stable
non-trivial asymmetric d.c.~states both become
time-dependent  via  Hopf bifurcations; one on each branch. The imperfection 
in the circuit is very small, so the loci of these bifurcations 
almost coincide and are marked 'Hopf' in Figure \ref{fig1}.  
The two asymmetric limit cycles which arise at the Hopf bifurcations appear to 
glue together leading to a large symmetric periodic orbit. This transition is 
denoted  by the line marked `Hom' in Figure \ref{fig1} and will be discussed in 
detail below. This symmetric limit cycle undergoes different types of bifurcation 
including symmetry-breaking and period doubling and may also become chaotic. 
Finally, within the oscillatory regime forward and reverse period doubling
sequences have been observed and these can be related to the Shil'nikov wiggle as
shown by Healey et al \cite{He}. The boundaries of this region are denoted by P2
in Figure \ref{fig1}.

\subsection{Imperfect gluing bifurcation}
\label{gluing0.6}

\begin{figure}
\includegraphics[width=8cm]{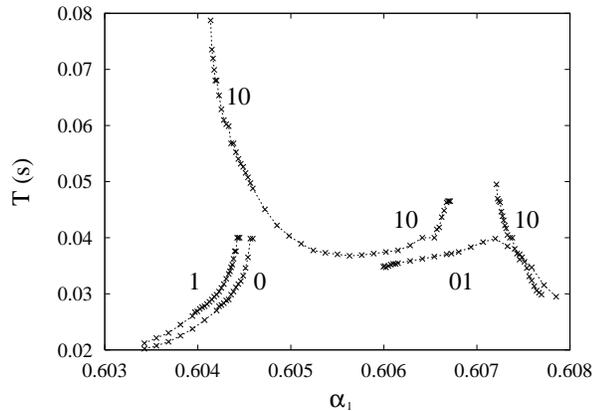}
\caption{Oscillation period of different periodic orbits at $\beta_1=0.6000$
plotted as a function of $\alpha_1$. `1' and `0' denote the orbits on the
asymmetric branches and `10', `01' are the glued orbits. }
\label{fig2}
\end{figure}

\begin{figure}
\includegraphics[width=6cm]{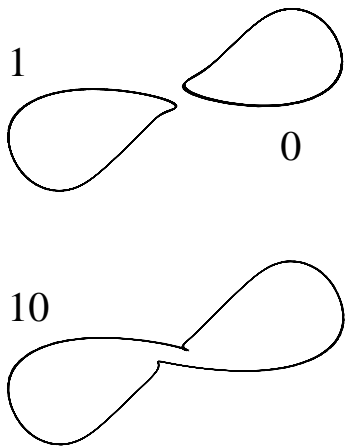}
\caption{Phase portraits of coexisiting asymmetric $(1, 0)$ and
symmetric $(10)$ periodic orbits at $\alpha_1=0.6041$ and $\beta_1=0.6000$}
\label{fig3}
\end{figure}

We first examine the influence of the imperfection on the gluing bifurcation
which occurs when the two asymmetric limit cycles join without
the presence of complicated dynamics. We chose $\beta_1$ sufficiently large 
($\beta_1\ge 0.59$ approximately) and $\alpha_1$ close to $\beta_1$ so
that the chaos which arises from period-doubling sequences on a Shil'nikov wiggle
is avoided and the dynamics is almost planar.  We present a
`typical' set of results for the orbit structure of the
oscillator in this regime in Figures \ref{fig2} and \ref{fig3} which were taken 
at $\beta_1=0.6000$. Figure \ref{fig2} shows the period of the various
simple orbits observed as a function of the parameter $\alpha_1$, and Figure
\ref{fig3} shows the form of the corresponding orbits -- the two small
asymmetric orbits are labelled by `1' and `0' respectively, and the
large amplitude orbit is labelled by `10', for reasons which will be
explained below.
 
If the electronic oscillator were symmetric then the development of the
orbits shown in Figure \ref{fig3} for $\alpha_1 = 0.6041$ would have a simple 
explanation in terms of gluing bifurcations \cite{CGT}: two periodic 
orbits which are the symmetric image of each other approach a stationary 
point and are `glued together' to form the single symmetric orbit with 
code `10'. At the bifurcation  the two smaller
periodic orbits touch at the stationary point, i.e. they are no longer periodic
(their period has diverged to infinity) and they form two homoclinic orbits,
biasymptotic to the stationary point.

As is clear from Figure \ref{fig2}, and as should be expected of a real physical
system, the oscillator is not perfectly symmetric. Hence it is
not surprising that the pair of homoclinic orbits which exist at a
single parameter value in the symmetric system seem to occur at different
parameter values in the oscillator. The results shown in
Figure \ref{fig2} also suggest that there is a third homoclinic bifurcation --
the bifurcation which creates the large amplitude `10' periodic orbit.

It can be seen in Figure \ref{fig2} that the period of 
both the small asymmetric orbits `1' and `0' increases as $\alpha_1$ increases
and they finally lose stability and  jump  to the `10'
orbit at $\alpha_{1}\approx 0.6045$ i.e. where the graphs of the variation of 
period are almost vertical. 
Moreover, the `0' orbit remains stable for slightly
higher values of $\alpha_1$ than the `1' orbit, emphasising that
the two orbits are not the images of each other under the symmetry. 
It should be noted that
the `1' orbit results from a Hopf bifurcation on the monotonic branch of
the disconnected pitchfork bifurcation. Therefore it loses stability before 
the `0' orbit. This is precisely what is predicted by the addition
of a constant term to the normal form. The
orbits shown in Figure \ref{fig3} all coexist at $\alpha_1=0.6041$ and are typical
examples of the limit cycles involved in this gluing bifurcation. The fact that
they can all coexist explains why hysteresis can be observed in the experiments.

There are three features in Figure \ref{fig2} which we will seek to explain
 theoretically in the next section: the break up of the gluing bifurcation,
 hysteresis, and also the extra bifurcations evident at larger values of
 $\alpha_1$. Before describing the theory we shall look at this latter
 sequence of bifurcations in more detail.

\subsection{Symmetry-breaking bifurcation of large periodic orbit}
\label{symbreak}
\begin{figure}
\center
\includegraphics[width=8cm]{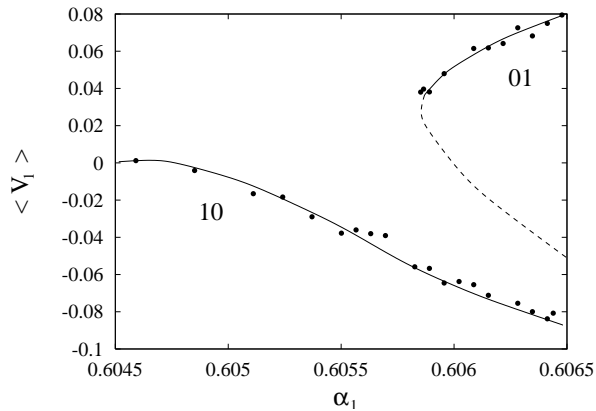}
\caption{Bifurcation diagram of symmetry-breaking bifurcation of
periodic orbits at $\beta_1=0.6000$. The mean of $V_1$ over $5000$ data points
is plotted.}
\label{fig4}
\end{figure}

\begin{figure}
\includegraphics[width=6cm]{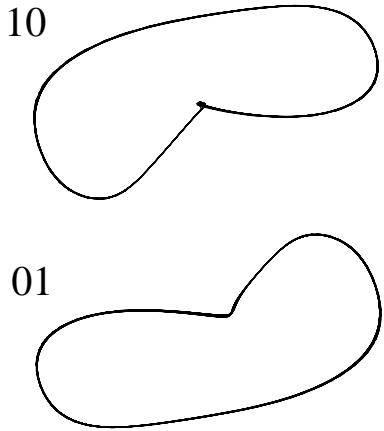}
\caption{Phase space portrait of coexisiting "large" periodic orbits $10$ and $01$
 at $\beta_1= 0.6000$ and $\alpha_1 = 0.6067$}
\label{fig5}
\end{figure}

It is known that symmetric systems cannot undergo period--doubling sequences
directly \cite{swift} but must first break their symmetry. Hence, we would expect
the large symmetric orbit formed by the gluing of the two asymmetric ones 
will suffer a symmetry breaking bifurcation, as predicted for the
symmetric Shil'nikov wiggle \cite{Gl}. This was observed 
at $\beta_1=0.6000$ for $\alpha_1$ just above $0.6059$.
The bifurcation was detected by measuring the mean voltage averaged over 
$150$ periods of the oscillation and plotting this as a function of 
$\alpha_1$. The resulting
bifurcation diagram is shown in Figure \ref{fig4} where we see that 
it has the form of a disconnected pitchfork. This diagram explains the
creation of the orbit labelled `01' in Figure \ref{fig2}. Note that
the original `10' orbit has a larger period but smaller $<V_1>$ than the newly 
created `01' orbit and so the branches in Figures \ref{fig2} and \ref{fig4} are
apparently reversed. 
Two typical asymmetric orbits on
respective branches are shown in Figure \ref{fig5} for $(\alpha_1, \beta_1) =(0.6067,06000)$. 
It is evident 
that the `10' orbit on the connected branch displays strong variation in period
for $\alpha_1>0.6065$ and then instability. However, the `01' orbit is virtually
constant over this range. Both orbits then show reduction in period for high
$\alpha_1$ values. Each orbit undergoes period doubling sequences to chaos
for $\alpha_1$ values greater than the range displayed in Figure \ref{fig2}.
The extra complications of period doubling and instability are topics for future
research.

\section{Theory}
\label{theory}

It is natural to think of the bifurcations observed in the system in terms
of two parameters. One of these, $\mu$ say, is the parameter of the (fictional)
symmetric system which has a gluing bifurcation as described in 
section \ref{gluing0.6}. The second parameter, $\epsilon$ say, is a measure
of how far the oscillator is from being perfectly symmetric, i.e. it is some
measure of imperfection with $\epsilon =0$ corresponding to the
perfectly symmetric system. Just as the standard imperfection theory for the
bifurcations of stationary points \cite{Sch} allows one to describe the
effect of asymmetry in terms of $\mu$ and $\epsilon$, our aim here is to give
an analogous description for general global bifurcations. We note that this
is in the spirit of the work of Glendinning \cite{gl} and Cox \cite{C} for the
particular case of Lorenz-like bifurcations.

\subsection{The basic picture}
\label{basics}
Suppose that $(\mu,\epsilon )=(0,0)$ denotes the point in parameter space at
which there are two symmetrically related homoclinic orbits. Consider either
one of these orbits. Since the existence of homoclinic orbits is codimension
one, there will
be a curve in parameter space through $(0,0)$ on which systems have a
homoclinic orbit which is a continuation of the given orbit. Thus, for typical
two-parameter families of systems, there will be two curves of homoclinic orbits
in parameter space, $G_0$ and $G_1$ say, which intersect at the origin and
which do not intersect the line $\epsilon =0$ again locally. The curve
$G_0$ (respectively $G_1$) is the locus of a homoclinic bifurcation which
creates or destroys the periodic orbit with code 0 (respectively, 1). The
one-parameter families of nearly symmetric systems such as the example
considered in the previous section would then correspond to some curve
in this two parameter space which has, for example, $\epsilon >0$ and
which passes close to $(\mu,\epsilon )=(0,0)$. Such a curve will intersect
both $G_0$ and $G_1$, but at different parameter values, so there will be
two simple homoclinic bifurcations at nearby parameter values on such a path.

The intersection of the loci of two homoclinic bifurcations (each to the same
stationary point) is a codimension two phenomenon which has been studied
by a number of authors\cite{CGT,Ga,GGT3,GGT4,GGT2,GGT1,Lyu,Tur}. The most important feature which all these bifurcations
have in common is that at least two other curves of homoclinic orbits emanate from
the intersection of $G_0$ and $G_1$, one in $\epsilon >0$, labelled $G_{10}$,
and the other in $\epsilon <0$ labelled $G_{01}$. The labelling describes
the order (in time) that the orbit passes through neighbourhoods of the basic
homoclinic orbits. These homoclinic orbits
are precisely the bifurcations needed to destroy or create (asymmetric)
periodic orbits with code `10' or `01'. Thus a typical path close to
$\epsilon =0$ will intersect $G_0$, $G_1$ and one of the curves $G_{01}$
or $G_{10}$. This explains the third homoclinic bifurcation observed in
Figure~\ref{fig2}. Roughly speaking, the difference between orbits created
by paths crossing $G_{10}$ and those created by crossing $G_{01}$ is
the difference between the orbits shown in Figure~\ref{fig5}.

\begin{figure}
\includegraphics[width=7cm]{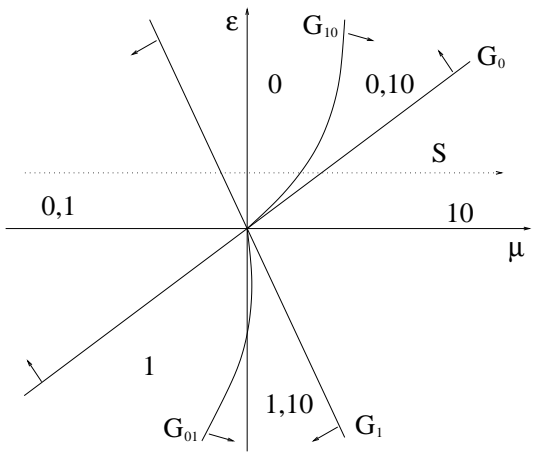}
\caption{The two parameter plane for the imperfect gluing bifurcation
in the planar case. A one parameter family of (imperfect) systems, $S$, is
indicated by a curve through the plane close to $\epsilon =0$. The arrows
indicate the direction in which orbits are created.}
\label{fig6}
\end{figure}

\begin{figure}
\includegraphics[width=7cm]{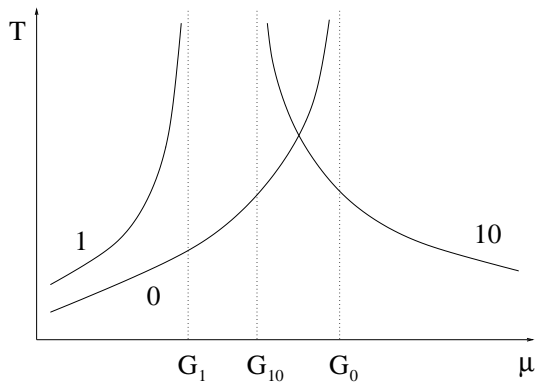}
\caption{Bifurcation diagram (period against parameter)
on the one-parameter path $S$ of Figure \ref{fig6}.}
\label{fig7}
\end{figure}

The details of the two-parameter bifurcation plane close to the intersection
of $G_0$ and $G_1$ depends upon the nature of the stationary point,
the configuration of the homoclinic orbits and a measure
of the amount of twisting of solutions about these orbits. The
nature of the stationary point is determined by the eigenvalues of the
Jacobian matrix of the flow which are closest to the imaginary axis. If,
up to complex conjugation, these are $\lambda_1$ and $\lambda_2$ with
${\rm Re}~\lambda_1<0<{\rm Re}~\lambda_2$ then the {\it saddle index},
$\delta$, defined by
\begin{equation}
\delta = -{\rm Re}~\lambda_1/{\rm Re}~\lambda_2
\end{equation}
plays an important role. The two-parameter
space near the intersection of $G_0$ and $G_1$ in the planar case is shown in 
Figure \ref{fig6} ($\lambda_1$ and $\lambda_2$ are real), where
the symmetry is a point symmetry about the stationary point  and the
direction of time may be chosen so that $\delta >1$. Each simple homoclinic
bifurcation creates a periodic orbit in the direction indicated by the arrow
on the bifurcation curve. The parameter plane is divided into six regions
by the curves of bifurcations, and the periodic orbits (from the local theory)
which exist in each region are indicated by their codes. The bifurcations
observed on the one-parameter path $S$ in Figure~\ref{fig6} are shown
in Figure~\ref{fig7}, which is the more conventional representation.

\subsection{Relationship with the experiment}
\label{tiein}
The curves sketched in Figure~\ref{fig7} are in reasonably good agreement with
the experimental ones in Figure~\ref{fig2}
except for the extra complications at larger parameter values described in
section~\ref{symbreak}. Also the fact, mentioned at the end of section
\ref{gluing0.6}, that all three of the orbits labelled `0', `1' and `10'
coexist for some values of $\alpha_1$. However, even these aspects can be incorporated
into our picture of imperfect global bifurcations. For smaller values
of $\beta_1$ Shil'nikov wiggles are observed, suggesting that
$\delta <1$ (and $\lambda_1$ is complex) in this parameter regime.
In this case,
as earlier, there may be symmetry-breaking and reverse
symmetry-breaking bifurcations of the symmetric orbit (in the perfectly
symmetric system) \cite{Gl}. The bifurcations observed in Figure~\ref{fig2} 
and described in more detail in Figure~\ref{fig4} are not
in the asymptotic region of applicability of the homoclinic theory
(large period, close to homoclinic bifurcation) and so we invoke
an extra pair of assumptions on the underlying symmetric system for our model: 
that
there is a symmetry-breaking and reverse-symmetry breaking bifurcation
on the symmetric orbit and that $\delta <1$. 

\begin{figure}
\includegraphics[width=7cm]{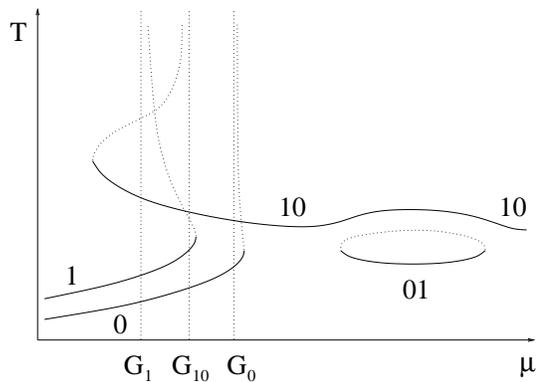}
\caption{Bifurcation diagram (period against parameter)
of the modified global bifurcation as suggested by Figure \ref{fig2}.}
\label{fig8}
\end{figure}

If $\delta <1$ then the curves
of homoclinic bifurcations are essentially as in Figure \ref{fig6} but the
direction of the bifurcations is reversed (more precisely, the diagram
is reflected about the $\epsilon -$axis) and the orbits created are saddles
(rather than stable, as would be the case if $\delta >1$). This now suggests the
new interpretation of Figure~\ref{fig2} which is shown in Figure~\ref{fig8}.
The major new feature is that since the orbits are created in the opposite
direction to the case with $\delta >1$ in Figure~\ref{fig6} and are unstable, the
points at which the orbits cannot be followed further ($\alpha_1 = 0.6041$ for the
`10' orbit and $\alpha_1 = 0.6025$ for the `0' and `1' orbits in Figure~\ref{fig2})
are now assumed to be saddle--node bifurcations. There are a number of
possible interpretations for the disconnected symmetry--breaking bifurcations, 
and one of these is shown in 
Figure~\ref{fig8}, although we make no claim that it is the
most likely. Note that the new arrangement of the homoclinic bifurcations
does provide a region of parameters where the orbits `0', `1', and `10' coexist and
are stable, as seen in the experiment.

The important feature of the analysis above is that two assumptions about  
the underlying mathematical model are sufficient to explain the orbits
observed in the experiment. It is worth emphasising that this can be done
{\it without} constructing the model equations explicitly, simply be
suggesting that any model equation must have various dynamical features.

\subsection{Other cases}
\label{more}
In the literature, codimension two global bifurcations
are generally described with $G_0$ and $G_1$ as the coordinate
axes of the bifurcation analysis. In this case the symmetric system may
be assumed to lie on the diagonal of the parameter space, with the asymmetry
parameter perpendicular to the diagonal (just tilt the
diagrams by $45^\circ$ to get an impression of the locus of bifurcations). 
It is, however, important to bear in mind that the curves $G_0$ and
$G_1$ appear to intersect with a very small angle of intersection in 
asymmetric perturbations of symmetric systems, whereas the standard analysis
depicts the intersection angle to be at right angles. Provided the intersection
is transversal the analysis is not affected, although it does mean that
the true picture for the asymmetric perturbation will be a very skewed 
version of the standard pictures.

The basic feature common to all the relevant types of bifurcation
we consider is that as the bifurcation parameter $\mu$ is varied, a
(more or less complicated) sequence of bifurcation is observed
with the net effect that a pair of periodic orbits (those we have
labelled `0' and `1') is destroyed, and a single large periodic orbit
is created. The precise details of the bifurcations depends on
the system, but it is still possible to make a number of general 
statements.

\subsubsection{The one-sided case}

If the direction of time can be chosen so that $\lambda_2$ is real 
and $\delta >1$
(cf. (1)) then the codimension one bifurcations on $G_0$ and $G_1$ are
one-sided and fairly general statements about the bifurcations involved 
{\it in the range of validity of the rigorous argument: large period and
parameters close to the intersection of $G_0$ and $G_1$} are
possible \cite{GGT1}. 
First, there are at most two periodic orbits, and second, any periodic orbit
has a very particular description in terms of the symbols `0' and `1' introduced
above. Technically, the sequences are rotation compatible sequences \cite{GGT1}, but
in practice a simple consequence is that periodic orbits have codes of the form
\begin{equation}
01^{n_1}01^{n_2}01^{n_3}01^{n_4}01^{n_5}\dots
\end{equation} 
where for all $i$, $n_i\in\{m,m+1\}$ for some $m>0$ (or the same with the roles of $0$
and $1$ exchanged). Moreover, the limit, $\rho$, of the number of $1s$ in the
sequence to the length of the sequence exists and is called the rotation number of
the orbit. In one case (the so-called stable 
orientable Lorenz-like case, see \cite{GGT2}), there is an infinite set of bifurcations along a typical path and at any one parameter after crossing the first
bifurcation curve, there is at most one periodic orbit. Moreover,
the rotation number varies continuously along the bifurcation path, implying the
existence of parameter values with non-periodic (but non-chaotic) attractors.  

If $\lambda_1$ is complex then the range of bifurcations is more complicated and
depends on the precise path taken through the parameter space. Here there are
regions of coexistence of certain periodic orbits -- those whose rotation numbers
${{p_1}\over{q_1}}$ and ${{p_2}\over{q_2}}$ are Farey neighbours, i.e.
$|p_1q_2-q_1p_2|=1$ -- but typical curves in parameter space do not intersect
most of these regions. A more complete list of the possibilities can
be found in \cite{Gl,Ga,GGT2}.

All the bifurcations of the rigorous analysis involve one-sided global 
bifurcations, and there are no local bifurcations on the branches of each 
periodic orbit. If these occur it is necessary to appeal to effects outside
the rigorous region of validity of the mathematical results -- this is made
much easier by an understanding of the two-sided bifurcations.

\subsubsection{The two-sided case: Shil'nikov's wiggle}

The symmetric bifurcation diagram of the Shil'nikov case ($\lambda_1$
complex, $\lambda_2$ real and $\delta <1$) is given in \cite{Gl}. The
locus of the pair of orbits (`0' and `1') in parameter-period space
oscillates as the period increases to infinity, with period-doubling
and reverse period-doubling bifurcations on every other branch. 
The symmetric orbit oscillates in a similar way, but with symmetry-breaking
bifurcations on every other branch. Breaking the symmetry of the system will
have two effects -- the global bifurcations which coincide in the
symmetric system will be split apart and the symmetry-breaking 
bifurcations will typically become
disconnected as described above. In the two--parameter diagram close
to the intersection of $G_0$ and $G_1$,
curves of more complicated bifurcations ($G_{01}$ and $G_{10}$) oscillate
rapidly and intersect each other (there are infinitely many other
curves of homoclinic bifurcations to complicate matters further).
For a typical asymmetric path there will be a single intersection with
$G_0$ and $G_1$, but potentially several intersections with $G_{10}$
and $G_{01}$.
The orbits created in the bifurcations involving $G_0$ and $G_1$ will 
lie on the usual Shil'nikov wiggle in the parameter--period plane as
observed experimentally (see Figure~\ref{fig9}).
The symmetric orbit, `10', can also be followed experimentally (see 
Figure~\ref{fig10}); there are multiple intersections
of the parameter path with $G_{10}$, i.e. extra bifurcations which create
and destroy the orbits labelled `10*'. Between the conjectured
intersection of the parameter path with $G_0$ and $G_{10}$ it is
possible to observe a stable orbit with code `100'. Such an orbit can 
be created from homoclinic orbits obtained from the gluing of the orbits
`10' and `0'. These bifurcations are expected due to the intersection of $G_0$ 
and $G_{10}$ in the two parameter analysis (cf. the $\delta >1$ case 
in \cite{GGT1}) which create extra curves of homoclinic orbits $G_{010}$
and $G_{100}$.  

\section{Experiment II}
\label{experiment2}

\begin{figure}
\includegraphics[width=8cm]{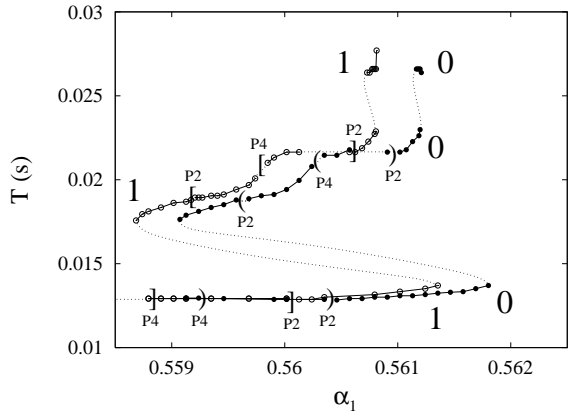}
\caption{Coexisiting Shil'nikov wiggle at $\beta=0.5317$. The branch noted with 
[ ] correspond to the `1'  and with ( ) to the `0' orbit respectively.}
\label{fig9}
\end{figure}

\begin{figure}
\includegraphics[width=8cm]{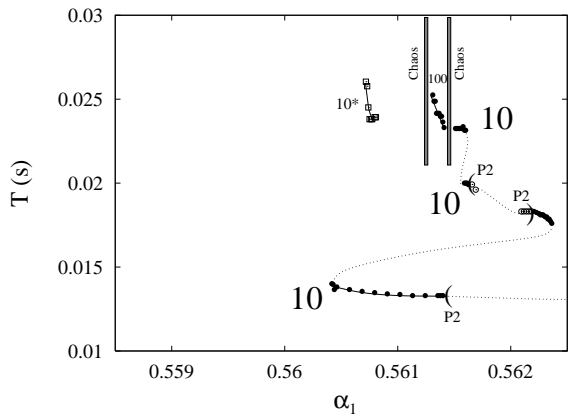}
\caption{Shil'nikov wiggle and gluing process of the `10' orbit at $\beta=0.5317$.
The period of the `10' orbits and the `100' orbit is rescaled by two and three 
respectively.}
\label{fig10}
\end{figure}

\begin{figure}
\center
\includegraphics[width=8cm]{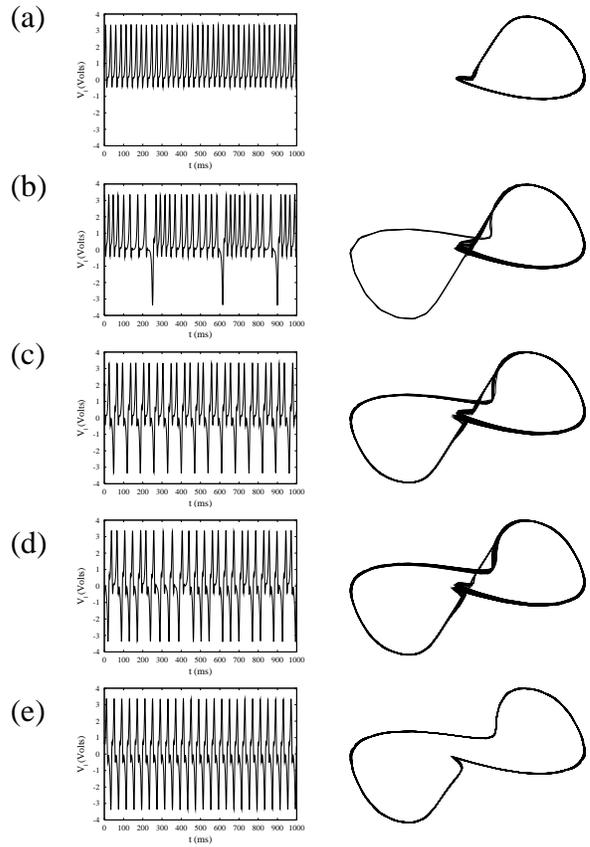}
\caption{Time series and phase portraits of different dynamical states
involved in imperfect gluing bifurcation at $\beta=0.5317$. 
(a) periodic orbit `0' on asymmetric branch at $\alpha_1=0.56119$, 
(b) chaos at $\alpha_1=0.56125$, 
(c) period-3 orbit `100' at $\alpha_1=0.56136$,
(d) chaos at $\alpha_1=0.56146 $, 
(e) symmetric periodic orbit `10' at $\alpha_1=0.56152$}
\label{fig11}
\end{figure}

The gluing process in the non--planar region of parameter space involves
complicated orbits which evolve on a Shil'nikov wiggle. A pair of these wiggles are
shown in Figure \ref{fig9} where the period is plotted as a function of
$\alpha_{1}$ at fixed $\beta_{1} = 0.5317$. Here the period of the orbit
approaches infinity through a sequence of folds where alternate branches are
unstable and indicated by dashed schematic lines in the figure. The stable
solutions undergo forward and reverse period--doubling sequences on the first two
folds whereas the highest period orbits only exist over a tiny range of the
parameter.In a perfectly symmetric system these two wiggles would overlap 
completely. The effect of the imperfection in the circuit is to displace the 
two curves from one another. 

A Shil'nikov wiggle has also been observed on the symmetric orbit and the
results are shown in Figure \ref{fig10}. There we can see three levels of the 
wiggle with period doubling sequences. The `10' orbits in this case were asymmetric but we
were unable to find the mirror image pairs of solutions in this case. We were,
however, able to observe them at smaller values of $\beta_1$. The gluing process
takes place on the third level with intervening sequences of chaos and a stable
`100' orbit; as expected from the discussion at the end of section \ref{more}. 
Note we also observed the `$10^*$' which
is an integral part of the gluing process as discussed in section \ref{more} above.
A set of time--series and phase portraits are displayed in Figure \ref{fig11}. 
The `0' orbit on the disconnected branch glues to the `10' large scale orbit
via two chaotic phases with an intermediate period--3 `100' sequence.

\section{Conclusion}

Although symmetric equations are frequently used to model almost symmetric
systems, we have shown that a more careful examination of experiments can
reveal features which do not appear in the symmetric models. In particular,
we have focussed here on global bifurcations which involve periodic
states of the system, and we have shown how a number of complicated bifurcation
diagrams observed in the experiments can be interpreted by appealing
to a theory of imperfect homoclinic bifurcations. 

A standard approach to the modelling of physical phenomena is to construct
a mathematical model of the experiment, and use this to either predict or explain
features of the experiment. This entails both the construction of the model and
the analysis of the model constructed. It is noticeable that in the approach
taken here we have appealed to properties of a model {\it without} having to
either construct or analyse the model. We have simply said that any mathematical
model of the experiments must have certain features, and that these features
lead to certain conclusions by the application of global bifurcation theory.

Bifurcation diagrams consistent with those of section~\ref{theory} have now
been observed in more physically interesting systems. Abshagen \cite{Abs} has
found bifurcation diagrams with a striking similarity to Figure~\ref{fig6}
in experimental data from fluid flow. We believe that the approach taken here
will find application in a broad variety of experiments in which symmetry,
or rather, almost-symmetry, plays a role.

\end{document}